\documentclass[twocolumn,pre,showpacs,amsmath,amssymb,floatfix]{revtex4}
\usepackage{graphicx}
\usepackage{dcolumn}

\begin{document}

\title{Effective free energy method for nematic liquid crystals in 
contact with structured substrates}

\author{L. Harnau$^{1,2}$, S. Kondrat$^{1,2}$, and A. Poniewierski$^3$}
\affiliation{
         $^1$ Max-Planck-Institut f\"ur Metallforschung,  
              Heisenbergstr.\ 3, D-70569 Stuttgart, Germany 
         \\
         $^2$ Institut f\"ur Theoretische und Angewandte Physik, 
              Universit\"at Stuttgart, Pfaffenwaldring 57, D-70569 Stuttgart, Germany
         \\
         $^3$ Institute of Physical Chemistry,
              Polish Academy of Sciences, Kasprzaka 44/52, 01-224 Warsaw, Poland}

\date{\today}

\begin{abstract}
We study the phase behavior of a nematic liquid crystal confined between a flat 
substrate with strong anchoring and a patterned substrate whose structure and 
local anchoring strength we vary. By first evaluating an effective surface free 
energy function characterizing the patterned substrate we derive an expression 
for the effective free energy of the confined nematic liquid crystal. Then we 
determine phase diagrams involving a homogeneous state in which the nematic director
is almost uniform and a hybrid aligned nematic state in which the orientation of 
the director varies through the cell. Direct minimization of the free energy 
functional were performed in order to test the predictions of the effective 
free energy method. We find remarkably good agreement between the phase 
boundaries calculated from the two approaches. In addition the effective energy 
method allows one to determine the energy barriers between two states in a 
bistable nematic device.
\end{abstract}

\maketitle

\section{Introduction}
The interest in anchoring phenomena and phenomena in confined nematic liquid 
crystals has largely been driven by their potential use in liquid crystal display 
devices. The  twisted nematic liquid crystal cell serves as an example. It consists 
of a nematic liquid crystal confined between two parallel walls, both providing 
homogeneous planar anchoring but with mutually perpendicular easy directions. In 
this case the orientation of the nematic director is tuned by the application of 
an external electric or magnetic field. A precise control of the surface alignment 
extending over large areas is decisive for the functioning of such devices.

Most studies have focused on nematic liquid crystals in contact with laterally 
uniform substrates. On the other hand substrate inhomogeneities arise rather 
naturally as a result of surface treatments such as rubbing. Thus the nematic 
texture near the surface is in fact non-uniform. This non-uniformity, however, 
is smeared out beyond a decay length proportional to the periodicity of the 
surface pattern. Very often the thickness of the non-uniform surface layer is 
considerably smaller than both the wavelength of visible light and the thickness 
of the nematic cell, i.e., the distance  between the two confining parallel walls. 
Hence optical properties of the nematic liquid crystal confined between such 
substrates correspond to those resulting from effective, uniform substrates. 

More recent developments have demonstrated that surfaces patterned with a large 
periodicity of some micrometers are of considerable interest from a technological 
point of view (see, e.g., Ref.~\cite{rasi:04} and references therein). 
A new generation of electro-optical devices relies  
on nematic liquid crystals with patterned orientation of the nematic director over 
small areas which can be achieved by chemically patterning the confining surfaces. 
For example, to produce flat-panel displays with wide viewing angles one can use pixels 
that are divided into sub-pixels, where each sub-pixel is defined by a different 
orientation of the nematic director, which is induced by the surface structure and
subsequently tuned by the electric field. In addition to the technological relevance, 
nematic liquid crystals in contact with non-uniform substrates provide the opportunity 
to study basic phenomena such as effective elastic forces between the substrates and 
phase transitions between various competing nematic textures (see, e.g., 
Ref.~\cite{harn:07} and references therein). 

Whereas the influence of homogeneous confining substrates on nematic liquid crystals 
is now well understood, the phase behavior of nematic liquid crystals in contact 
with chemically or geometrically patterned substrates is still debated. One might 
suppose that theoretical 
calculations based on continuum theories should resolve the properties of 
nematic liquid crystals in contact with patterned substrates
\cite{berr:72,faet:87,barb:92,bari:96,card:96,qian:96,qian:97,brya:98,four:99a,four:99b,moce:99,barb:99,brow:00,kond:01,wen:02,wen:02a,patr:02,kond:03,zhan:03,poni:04,tsui:04,harn:04,bata:05,kond:05,desm:05,harn:05,wan:05,kiyo:06,yeun:06,athe:06,bech:06}. However, 
such calculations are numerically demanding because two- or
three-dimensional  grids have to be used because of the broken symmetry
due to the surface pattern.  Moreover, it is very challenging to
determine metastable states and energy barriers  between them which are
important for the understanding of bistable nematic devices
\cite{boyd:80,chen:82,dozo:97,barb:97,bock:98,zhua:99,guo:00,denn:01,kim:01,kim:02,davi:02,lee:03,parr:03,brow:04,parr:04,barm:04,sikh:05,mizo:05,parr:05,uche:05,harn:06,barm:06,uche:06,schm:06}.
In the present paper we adopt a different strategy which takes the
advantage of the finite decay length characterizing the influence of the
surface pattern on the  nematic liquid crystal in the direction
perpendicular to the substrate. We determine  first an anchoring energy
function and an average surface director orientation  of the patterned
substrate and obtain an effective free energy for the nematic liquid 
crystal cell under consideration. We find remarkably good agreement
between the  phase diagrams of various systems calculated using this
effective  free energy function on the one hand and the original free
energy  functional on the other hand.

\section{Effective free energy function }
\subsection{Continuum theory}
The continuum theory for liquid crystals has its origin dating back to at least 
the work of Oseen \cite{osee:25} and Zocher \cite{zoch:25}. This early version of 
the continuum theory for nematic liquid crystals played an important role for the
further development of the static theory and its more direct formulation by Frank
\cite{fran:58}. The Frank theory is formulated in terms of the so-called 
nematic director $\hat{{\bf n}}=\hat{{\bf n}}({\bf r})$, $|\hat{{\bf n}}|=1$, and 
its possible spatial distortions. The nematic director describes the direction of 
the locally averaged molecular alignment in liquid crystals. In a nematic liquid 
crystal the centers of mass of the liquid crystal molecules do not exhibit 
long-ranged order. The molecules can translate freely while being aligned, 
on average, parallel to one another and to the nematic director. 
It is known that if an initially uniform nematic liquid crystal is distorted by 
external forces, it relaxes back to the uniform state after the disturbing influence 
is switched off, signaling that the uniform configuration represents a thermodynamically 
stable state. Therefore it is assumed that there is a cost in free energy associated 
with elastic distortions of the nematic director of the form 
\begin{eqnarray} 
F_{elas}[\hat{\bf n}({\bf r)}]&=&\frac{1}{2}\int_V d^3
r\,\left[K_{11}\left(\nabla\cdot\hat{{\bf n}}\right)^2\right.\nonumber
\\&+&\!\!\!\left.K_{22}(\hat{{\bf n}}\cdot(\nabla \times\hat{{\bf n}}))^2
+K_{33}(\hat{{\bf n}}\times(\nabla \times \hat{{\bf n}}))^2\right],\nonumber
\\&&
\label{eq1}
\end{eqnarray}
where $V$ is the volume accessible to the nematic liquid crystal and
$K_{11}$, $K_{22}$,  and $K_{33}$ are elastic constants associated with
splay, twist, and bend distortions,  respectively. The elastic constants 
depend on temperature and are commonly of the order $10^{-12}$ to
\mbox{$10^{-11}\,\mathrm{N}$}. Sometimes, for example, when the relative
values of the elastic constants are unknown  or when the resulting
Euler-Lagrange equations are complicated, the one-constant 
approximation $K=K_{11}=K_{22}=K_{33}$ is made. In
this case the  elastic free energy functional reduces to
\begin{eqnarray} \label{eq2}
F_{elas}[\hat{\bf n}({\bf r)}]&=&\frac{K}{2}\int_V d^3 r\,\left(\nabla\hat{{\bf n}}\right)^2\,.
\end{eqnarray}
In the presence of surfaces the bulk free energy $F_b=F_{elas}$ must be supplemented by 
the surface free energy $F_s$ such that the total free energy is given by $F=F_b+F_s$.
In the corresponding equilibrium Euler-Lagrange equations $\delta F/\delta \hat{{\bf n}}=0$,
$F_s$ leads to appropriate boundary conditions. The description of the nematic director
alignment at the surfaces forming the boundaries is called anchoring. In addition to the
so-called free boundary condition where there is no anchoring, one considers weak and 
strong anchoring. If there are no anchoring conditions imposed on $\hat{{\bf n}}$ at the
boundary, the bulk free energy $F_b$ is minimized using standard techniques of the calculus 
of variations. In the case of strong anchoring it is also sufficient to minimize 
the bulk free energy but subject to $\hat{{\bf n}}$ taking prescribed fixed values 
at the boundary. In the case of weak anchoring the total free energy $F$, which includes 
the surface free energy $F_s$, has to be minimized. The most commonly used expression for 
the surface free energy is of the form proposed by Rapini and Papoular by \cite{rapi:69}:
\begin{equation} \label{eq3}
F_s[\hat{\bf n}({\bf r)}]=\frac{1}{2}\int_{S} d^2 r\,w({\bf r}) (\hat{{\bf n}}\cdot{\nu})^2.
\end{equation} 
The integral runs over the boundary and $w=w({\bf r})$ is the corresponding anchoring 
strength that characterizes the surface. The local unit vector perpendicular to the surface 
is denoted as $\hat{{\bf n}}$. For negative $w$, this contribution favors an orientation of 
the molecules perpendicular to the surface, while positive $w$ favor degenerate planar 
orientations at the surface. The absolute value of the anchoring strength is commonly of 
the order $10^{-6}$ to \mbox{$10^{-2}$ N/m}.

\subsection{The model}
Here we consider a nematic liquid crystal confined between a patterned substrate at 
$z=0$ and a flat substrate at $z=D$ where the $z$ axis is normal to the flat substrate. 
As Figs.~\ref{fig1} (a), (c), and (d) illustrate, the lower substrate is
characterized by geometrical and/or chemical patterns of periodicity $p$ along 
the $x$ axis. Moreover, the system is translationally invariant in the $y$ direction.
Within the one-constant approximation [Eq.~(\ref{eq2})] the total free energy functional
is given by
\begin{eqnarray} 
\lefteqn{F[\theta(x,z);\theta_D,D]=}\nonumber
\\&&\frac{KL}{2}\int\limits_0^p dx\,
\int\limits_{z_0(x)}^D dz\, [\nabla\theta(x,z)]^2+F_s[\theta(x,z_0(x))]\,,
\label{eq4}
\end{eqnarray}
where $L$ is the extension of the system in $y$ direction, $z_0(x)$ is the 
surface profile of the patterned substrate, and ${\hat{\bf n}}=(\sin\theta,0,\cos\theta)$.
At the upper surface strong anchoring $\theta(x,z=D)=\theta_D$ is imposed.
Twist is not considered, i.e., $K_{22}=0$. Of course, the analysis can be 
straightforwardly extended to the case of different splay and bend constants 
$K_{11}$ and $K_{33}$, respectively, but this aspect of the problem is important 
only if the analysis is supposed to yield quantitative results for a specific 
nematic liquid crystal. The surface contribution $F_s[\theta(x,z_0(x))]$ includes 
the anchoring energy for which the Rapini-Papoular form [Eq.~(\ref{eq3})] is 
adopted:
\begin{figure}[t!]
\includegraphics[width=8cm]{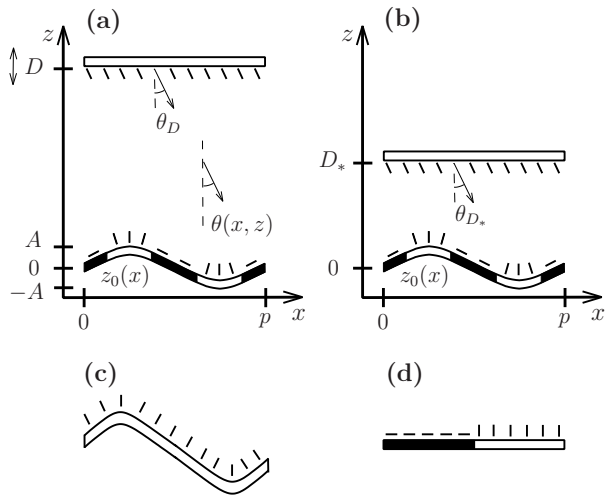} 
\caption{(a) The system under consideration consists of a nematic liquid crystal 
confined between two substrates at a mean distance $D$. The upper flat substrate 
induces strong anchoring, i.e., $\theta(x,z=D)=\theta_D$, while the lower substrate 
is characterized by a surface pattern of period $p$ in $x$ direction. The system is
translational invariant in $y$ direction perpendicular to the plane of the figure.
In (a) the lower sinusoidally grating surface (groove depth $A$) is endowed with an
alternating stripe pattern  of locally homeotropic anchoring (white bars) and homogeneous
planar anchoring (black bars). The period of the chemical pattern is half the period of 
the surface grating $p$. In (c) and (d) a pure geometrically structured lower 
substrate and pure chemically patterned lower substrate, respectively, are shown.
The anchoring direction at the substrates is schematically represented by black rods.
Quantitatively reliable predictions of the phase behavior of a nematic liquid crystal 
confined between two substrates at an {\it arbitrary} mean distance $D$ in (a) can 
be achieved if the effective free energy function [Eqs.~(\ref{eq6}) - (\ref{eq8})] is 
analyzed. To this end the free energy functional of the nematic liquid crystal 
confined between two substrates at a {\it single} and rather small mean distance $D_*$
in (b) has to be minimized as is discussed in the main text.}
\label{fig1}
\end{figure}
\begin{eqnarray} \label{eq5}
\lefteqn{F_s[\theta(x,z_0(x))]=}\nonumber
\\&&\!\!\!\!\!\frac{L}{2}\int^p_0 dx\,
w(x)\frac{\left(-\sin(\theta_0(x))z'_0(x)+\cos(\theta_0(x))\right)^2}
{\sqrt{1+(z'_0(x))^2}},
\end{eqnarray} 
where $\theta_0(x)\equiv\theta(x,z=z_0(x))$. Equations (\ref{eq4}) and (\ref{eq5}) 
together with the boundary condition for the nematic director at the upper surface 
completely specify the free energy functional for the system under consideration.
The Euler-Langrange equations resulting from the stationary conditions of the 
total free energy with respect to the nematic director field can be solved numerically 
on a sufficiently fine two-dimensional grid using an iterative method. In order to 
obtain both stable and metastable configurations, different types of initial 
configurations are used. However, due to the pattern of the lower surface the 
determination of the director field and the phase diagram turns out to be a 
challenging numerical problem in particular in the case of large cell widths 
$D$. Moreover, the energy barrier between two metastable states cannot be determined 
this way.

\subsection{Effective free energy function}
Here we map the free energy functional $F[\theta(x,z);\theta_D,D]$ [Eq.~(\ref{eq4})]
of a nematic liquid crystal cell with {\it arbitrary} width $D$ and 
arbitrary anchoring angle $\theta_D$ at the upper surface (see Fig.~\ref{fig1}~(a))
onto the effective free energy function
\begin{eqnarray} \label{eq6}
F^{(eff)}(\tilde{\theta}_0,\theta_D,D)&\!\!=\!\!&\frac{KLp}{2D}
\left(\theta_D-\tilde{\theta}_0\right)^2\!+\!F_s^{(eff)}(\tilde{\theta}_0),
\end{eqnarray}
where the average surface director orientation $\tilde{\theta}_0$ \cite{poni:97}
at the lower patterned surface is given by 
\begin{eqnarray} \label{eq7}
\tilde{\theta}_0(\theta_{D_*},D_*)&=&
\theta_{D_*}-\frac{D_*}{KLp}\frac{\partial}{\partial\theta_{D_*}}
F(\theta_{D_*},D_*)|_{min}\,.
\end{eqnarray}
The effective surface free energy function $F_s^{(eff)}$ characterizing the 
anchoring energy at the patterned surface can be written as 
\begin{eqnarray} \label{eq8}
\lefteqn{F_s^{(eff)}(\tilde{\theta}_0(\theta_{D_*},D_*),D_*)=}\nonumber
\\&& F(\theta_{D_*},D_*)|_{min}
-\frac{KLp}{2D_*}\left(\theta_{D_*}-\tilde{\theta}_0(\theta_{D_*},D_*)\right)^2.\nonumber
\\&&
\end{eqnarray}
In order to calculate $\tilde{\theta}_0$ and $F_s^{(eff)}$
explicitly, we first determine numerically the minimum of the free
energy $F(\theta_{D_*},D_*)|_{min}$ of the nematic liquid crystal cell
for  a {\it single} and rather small value $D=D_*$ and arbitrary
anchoring angle  $\theta_{D_*}$ at the upper surface (see
Fig.~\ref{fig1}~(b)). Thereafter the  phase behavior, energy barriers
between metastable states, and effective anchoring  angles for the
system of interest (Fig.~\ref{fig1}~(a)) can be obtained for  {\it
arbitrary} values of $D$ and $\theta_D$ from
$F^{(eff)}(\tilde{\theta}_0,\theta_D,D)$ as a function of the single
variable $\tilde{\theta}_0$. Such a calculation is considerably less
challenging than minimizing the original free energy functional 
$F[\theta(x,z);\theta_D,D]$ with respect to $\theta(x,z)$ on a
two-dimensional  $(x, z)$ grid. However, the effective free energy
method is applicable only if Eq.~(\ref{eq7}) can be inverted in order to
obtain $\theta_{D_*}(\tilde{\theta}_0)$ which is needed as input into
Eq.~(\ref{eq8}). The condition for this inversion  follows from
Eq.~(\ref{eq7}): 
\begin{eqnarray} \label{eq9}
\left(1-\frac{D_*}{KLp}\frac{\partial^2}{\partial\theta^2_{D_*}}
F(\theta_{D_*},D_*)|_{min}\right)^2>0\,.
\end{eqnarray}
Moreover, $F_s^{(eff)}(\tilde{\theta}_0(\theta_{D_*},D_*),D_*)$ is practically independent 
of the cell width $D_*$ provided $D_* \gtrsim p$ implying that the interfacial region 
above the lower substrate does not extend to the upper substrate.

Before studying the nematic liquid crystal in contact with the patterned substrates 
shown in Fig.~\ref{fig1} it is instructive to analyze first the nematic liquid crystal
confined between two homogeneous flat substrates at a distance $D$. The upper surface 
induces strong anchoring, i.e., \mbox{$\theta(z=D)=\theta_D$}. The free energy 
functional defined in Eq.~(\ref{eq4}) follows as 
\begin{eqnarray} \label{eq10}
F[\theta(z);\theta_D,D]&\!\!=\!\!&\frac{KLp}{2}\int\limits_0^D
dz\, \left(\frac{d \theta(z)}{d z}\right)^2+F_s(\theta_0).
\end{eqnarray}
The solution of the Euler-Langrange equation $\partial^2_z\theta(z)=0$ 
subject to the boundary condition at the upper surface $\theta(z=D)=\theta_D$
interpolates linearly between the top and bottom surfaces:

\begin{eqnarray} \label{eq11}
\theta(z)&=&\theta_D-\frac{1}{D}(D-z)(\theta_D-\theta_0)\,,
\end{eqnarray}
where $\theta(z=0)=\theta_0$ follows from the boundary condition at the 
lower surface. With this solution of the Euler-Lagrange equation the 
minimized free energy function reads
\begin{eqnarray} \label{eq12}
F(\theta_{D_*},D_*)|_{min}&=&
\frac{KLp}{2D_*}\left(\theta_{D_*}-\theta_0\right)^2+F_s(\theta_0)\,.\nonumber
\\&&
\end{eqnarray}
It follows directly from Eqs.~(\ref{eq6}) - (\ref{eq8}) that 
$\tilde{\theta}_0=\theta_0$, $F_s^{(eff)}=F_s$, and
\begin{eqnarray} \label{eq13}
F^{(eff)}(\theta_0;\theta_D,D)&=&\frac{KLp}{2D}
\left(\theta_D-\theta_0\right)^2+F_s(\theta_0). 
\end{eqnarray}
Hence in the case of homogeneous confining substrates the effective free energy 
function [Eq.~(\ref{eq13})] agrees exactly with the minimized free energy function
of the original system [Eq.~(\ref{eq12})].

\section{Applications}

\subsection{Geometrically and chemically patterned substrates}
\label{IIIA}

\begin{figure}[t!]
\includegraphics[width=8cm]{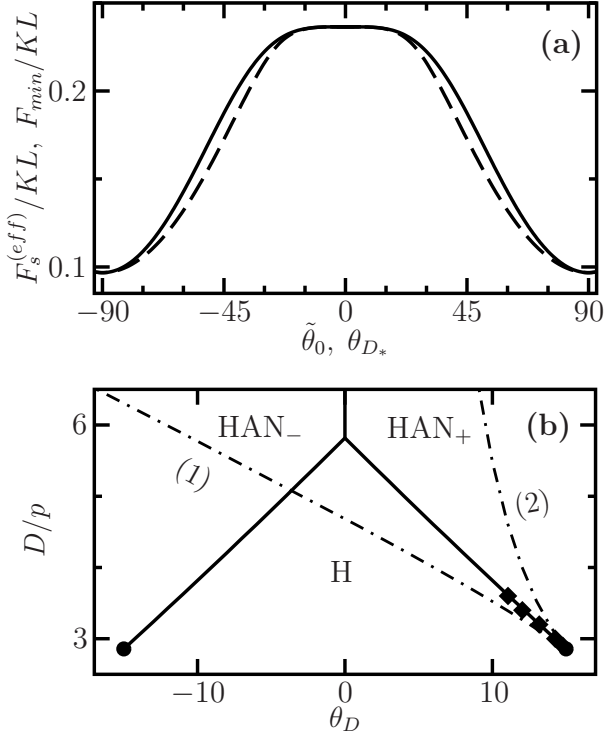} 
\caption{(a) The minimized free energy \mbox{$F(\theta_{D_*},D_*=p)|_{min}$} (dashed line) 
and the effective surface free energy
\mbox{$F_s^{(eff)}(\tilde{\theta}_0,D_*=p)$} (solid line) of a nematic
liquid crystal confined between a flat surface with strong anchoring 
and a chemically patterned sinusoidal surface with groove depth
$A/p=0.09$  (see Fig.~\ref{fig1}~(b)). $L$ is the extension of the cell
in the invariant  $y$ direction, $K$ is the isotropic elastic constant,
and the anchoring strength on  the homeotropic stripes (white bars in
Fig.~\ref{fig1}~(b)) and planar anchoring  stripes (black bars in
Fig.~\ref{fig1}~(b)) are $pw_H/K=1$ and $pw_P/K=2.5$, respectively. (b)
Phase diagram of the same system as a function of the anchoring  angle
at the upper flat substrate $\theta_D$ and the cell width $D$ (see
Fig.~\ref{fig1}~(a)). The solid line denotes first order phase
transitions between a homeotropic (H) and  hybrid aligned nematic
(HAN$_+$ and HAN$_-$) phases. At $\theta_D = 0$ and $D/p \approx 5.8$
there is a triple point where the HAN$_+$, HAN$_-$, and H states coexist.
The solid circle marks the critical point at  $D_{cr}/p \approx 3$ and
$\theta_D^{(cr)} \approx \pm 15^{\circ}$. The limits of metastability of 
the HAN$_+$ (1) and the H (2) state are denoted by the dot-dashed lines.
The limit of metastability of the H state for $\theta_D < 0$ and the
HAN$_-$ state are not shown for clearness. The lines and the solid
circle follow from analyzing the effective free energy function
[Eqs.~(\ref{eq6}) - (\ref{eq8})] while the diamonds represent the phase
boundary of thermal equilibrium as obtained from a direct minimization
of the underlying free  energy functional [Eqs.~(\ref{eq4}) and
(\ref{eq5})].}
\label{fig2}
\end{figure}

In the previous section we have shown that we can describe a nematic liquid
crystal  confined between two homogeneous substrates by the effective
free energy function [Eq.~(\ref{eq6})]. In this subsection we apply this
approach to the particular case of a nematic  liquid crystal confined
between a chemically patterned sinusoidal surface and a flat substrate
with strong homeotropic anchoring (see Fig.~\ref{fig1}~(a)). The 
surface profile of the grating surface is given by $z_0(x)=A\sin \left(q
x\right)$,  where $A$ is the groove depth and $p=2\pi/q$ is the period.
As Figure \ref{fig1} (a)  illustrates, the surface exhibits a pattern
consisting of alternating stripes   with locally homeotropic and
homogeneous planar anchoring. The projection of  the widths of the
stripes onto the $x$ axis is $p/4$ and the anchoring strength is
specified by a periodic step function: $w(x)=-w_H$ and $w_P$ for values
of $x$  on the homeotropic and planar stripes, respectively. Figure
\ref{fig2} (a) displays $F(\theta_{D_*},D_*=p)|_{min}$ (dashed line) and 
$F_s^{(eff)}(\tilde{\theta}_0,D_*=p)$ (solid line) for $pw_H/K=1$, 
$pw_P/K=2.5$, and $A/p=0.09$. The shapes of
$F(\theta_{D_*},D_*=p)|_{min}$ as a function of $\theta_{D_*}$ and
$F_s^{(eff)}(\tilde{\theta}_0,D_*=p)$ as a function of
$\tilde{\theta}_0$ are  rather similar because $\tilde{\theta}_0\approx
\theta_{D_*}$ (see Eq.~(\ref{eq8})) for this set of model parameters.
Figure \ref{fig2} (b) displays the phase  diagram plotted as a function
of the anchoring angle $\theta_D$ at the upper substrate and the mean
separation of the substrates $D$. The calculations demonstrate the
existence  of two (stable or metastable) nematic director
configurations: the homeotropic (H) phase, in which the director field
is almost uniform and parallel to the anchoring  direction imposed at
the upper surface, i.e.,  ${\hat{\bf n}}_H = (\sin\theta_D, 0,
\cos\theta_D)$, and the hybrid aligned nematic (HAN) phase,  in which
the director field varies from ${\hat{\bf n}}_H$ at the upper surface to
nearly planar orientation through the cell. Note that there are two HAN
textures: HAN$_+$ and HAN$_-$ corresponding to positive and negative
average surface angles at the lower surface (see also
Fig.~\ref{fig:diagram} below). For small anchoring angles $\theta_D$ the 
HAN phases are stable provided the cell width is larger than $D_{coex}$
(more precisely, the HAN$_+$ texture is stable for $\theta_D > 0$ while
the HAN$_-$ texture is stable for $\theta_D < 0$ and they coexist at $\theta_D =
0$). For smaller distances  between the substrates $D<D_{coex}$ the HAN
phases are no longer stable because distortions  of the director field are
too costly in the presence of the dominating strong anchoring  at the
upper surface. The comparison of the phase boundary of thermal
equilibrium as obtained from the effective free energy method
[Eqs.~(\ref{eq6}) - (\ref{eq8}), and  solid line in Fig.~\ref{fig2}~(b)]
and the direct minimization of the underlying free energy functional
[Eqs.~(\ref{eq4}) and (\ref{eq5}), and diamonds in Fig.~\ref{fig2}~(b)]
demonstrate  the reliability of the effective free energy method.

\begin{figure}[t!]
\includegraphics[width=8cm]{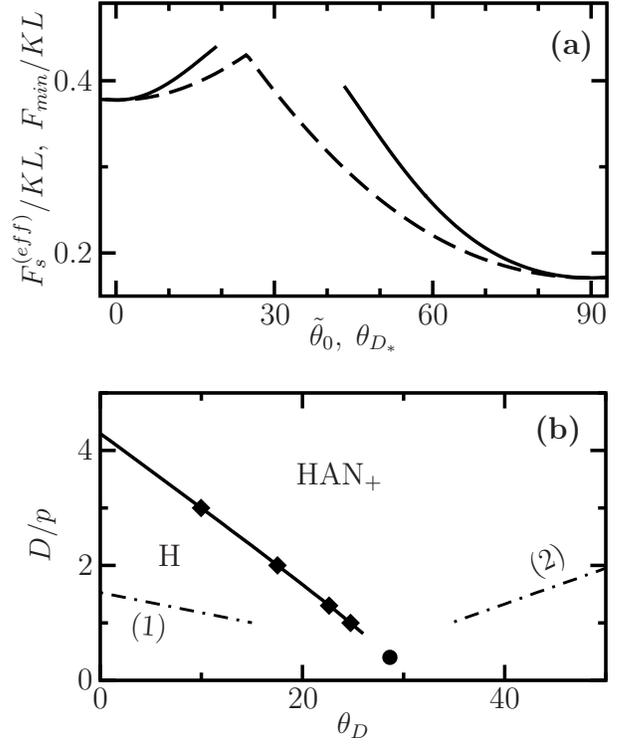} 
\caption{(a) The minimized free energy \mbox{$F(\theta_{D_*},D_*=p)|_{min}$} (dashed line) 
and the effective surface free energy
\mbox{$F_s^{(eff)}(\tilde{\theta}_0,D_*=p)$} (solid line) of a nematic
liquid crystal confined between a flat surface with strong anchoring 
and a chemically patterned sinusoidal surface with groove depth
$A/p=0.09$ (see Fig.~\ref{fig1}~(b)). The anchoring strength on the
homeotropic stripes and planar  anchoring stripes are $pw_H/K=2.5$ and
$pw_P/K=6$, respectively. (b) Phase diagram of the same system as a
function of the anchoring  angle at the upper flat substrate $\theta_D$
and the cell width $D$ (see Fig.~\ref{fig1}~(a)). The solid line denotes
the first order phase transition between a homogeneous (H) and  hybrid
aligned nematic (HAN$_+$) phase. The solid circle marks the critical
point at  $D_{cr}/p \approx 0.4$ and $\theta_D^{(cr)} \approx
29^{\circ}$. The limits of metastability of  the HAN$_+$ (1) and the H
(2) state are denoted by the dot-dashed lines. The lines follow from
analyzing the effective free energy function [Eqs.~(\ref{eq6}) -
(\ref{eq8})] while the diamonds and the solid circle represent the phase
boundary and a critical point as obtained from a direct minimization of
the underlying free energy functional [Eqs.~(\ref{eq4}) and
(\ref{eq5})]. For $D \lesssim p$ the phase transition as well as the
limits of metastability cannot be determined using the effective free
energy function because $F_s^{(eff)}(\tilde\theta_0)$ is not known in
the region close to its maximum. For clearness, only the phase diagram
for positive $\theta_D$ is shown (c.f. Fig.~\ref{fig2}).}
\label{fig3}
\end{figure}

We note that the phase transition between the H and HAN textures is first
order despite the fact that the effective 
surface free energy favors monostable planar anchoring, i.e., $F_s^{(eff)}$ exhibits only 
a minimum at $\tilde{\theta}_0=\pi/2$ in the interval $\tilde{\theta}_0 \in [0,\pi/2]$.
A first order phase transition in a nematic liquid crystal device with a monostable 
anchoring condition on a homogeneous lower substrate has been predicted for the special 
case $\theta_D=0$ in Refs.~\cite{parr:03,parr:04} using the empirical expression
\begin{eqnarray} \label{eq14}
F_s(\theta_0)&=&\frac{w_0}{2}\sin^2(2\theta_0)+w_1\sin^2(\theta_0)
\end{eqnarray}
as input into Eq.~(\ref{eq13}). For $w_1<2w_0$ this surface free energy has two 
minima at $\theta_0=0$ and $\theta=\pi/2$ in the interval $\theta_0 
\in [0,\pi/2]$, and as such is bistable, while for $w_1>2w_0$, only the minimum
$\theta_0=0$ exists, i.e., the surface is monostable. To study the stability limit of 
the H phase we expand $F^{(eff)}(\theta_0)$ in Eq.~(\ref{eq13}) around $\theta_0=0$ up to sixth order:
\begin{eqnarray} \label{eq15}
F^{(eff)}(\theta_0) &\approx& F_s(0)+\frac{1}{2}\left(\frac{KLp}{D}+F_s^{''}(0)\right)\theta_0^2 \nonumber 
\\ && + \frac{1}{4!}F_s^{(4)}(0)\theta_0^4+\frac{1}{6!}F_s^{(6)}(0)\theta_0^6\,.
\end{eqnarray}
The H phase corresponds to a local minimum of $F^{(eff)}(\theta_0)$ in 
Eq.~(\ref{eq13}) if $D<D_{cr}$, where
\begin{eqnarray} \label{eq16}
D_{cr}=-\frac{KLp}{F^{''}_s(0)}\,.
\end{eqnarray}
A standard bifurcation analysis reveals that the transition from the H phase to the 
HAN phase can be either first order or continuous. The transition is continuous if 
$F_s^{(4)}(0)>0$, first order if $F_s^{(4)}(0)<0$, and $F_s^{(4)}(0)=0$ corresponds 
to a tricritical point. In the case of a first order phase transition the phase 
boundary of thermal equilibrium is given by 
\begin{eqnarray} \label{eq17}
\frac{KLp}{D_{tr}}=\frac{10\left(F_s^{(4)}(0)\right)^2}{16 F_s^{(6)}(0)}-F_s^{''}(0)\,.
\end{eqnarray}
We emphasize that the order of the phase transition depends only on the surface free 
energy $F_s(\theta_0)$ close to $\theta_0=0$ for $\theta_D=0$. Therefore it is possible 
to have a first order phase transition even with a monostable surface characterized by 
a monotonic surface free energy such as the one shown in Fig.~\ref{fig2}~(a) for 
${\tilde \theta}_0\in [0,\pi/2]$ or the empirical equation (\ref{eq14}) with $w_1>2w_0$
as well the more general expression
\cite{four:99a}
\begin{eqnarray} \label{eq18}
F_s(\theta_0)&=&\sum\limits_{n=0}^\infty\left[a_n \cos(2n\theta_0)
+b_n \sin(2n\theta_0)\right]
\end{eqnarray}
with appropriate parameters $a_n$ and $b_n$. 

First order phase transitions between the 
H and HAN texture are of particular interest for bistable liquid crystal displays. 
In a bistable liquid crystal display the two molecular configurations corresponding 
to light and dark states are locally stable in the thermodynamic space when the applied  
voltage is removed \cite{davi:02,harn:06}. Therefore, power is needed only to switch 
from one stable state to another, in contrast to monostable liquid crystal displays which 
require power to switch and to maintain the light and the dark states. 

We now turn our attention to the case that it is not possible to evaluate 
$\theta_{D_*}(\tilde{\theta}_0,D_*)$ from $\tilde{\theta}_0(\theta_{D_*},D_*)$
[Eq.~(\ref{eq7})] because the condition for this inversion is not satisfied 
[Eq.~(\ref{eq9})]. To this end we have chosen the parameters $pw_H/K=2.5$,  
$pw_P/K=6$, and $A/p=0.09$ for the system shown in Fig.~\ref{fig1}~(a). 
Figure \ref{fig3} (a) display \mbox{$F(\theta_{D_*},D_*=p)|_{min}$} (dashed line) 
and \mbox{$F_s^{(eff)}(\tilde{\theta}_0,D_*=p)$} (solid line) while the corresponding
phase diagram is shown in Fig.~\ref{fig3}~(b). As is apparent from the solid line
in Fig.~\ref{fig3}~(a) it is not possible to determine the effective surface 
free energy function for the all values of $\tilde{\theta}_0$ because the upper 
flat substrate at $D_*$ is too far away from the lower patterned substrate in 
order to induce all possible average anchoring orientations $\tilde{\theta}_0$.
In other words, the anchoring energy at the patterned substrate is too large 
to be balanced by the elastic energy for the chosen mean distance $D_*$ between 
the substrates (see Fig.~\ref{fig1}~(b)). Nevertheless Fig.~\ref{fig3}~(b)
demonstrates that even this partial information about the effective surface free 
energy function can be used to calculate the phase diagram for cell widths 
sufficiently larger than the width at the critical point $D_{cr}$.

\subsection{Purely geometrically structured substrates}
\label{IIIB}

In the last subsection we have shown that with a suitable chemical and
geometrical surface morphology on one of the interior surfaces of a
liquid crystal cell, two stable  nematic director configurations can be
supported. The zenithally bistable nematic devices that have been
studied recently \cite{parr:03,brow:04,parr:04,parr:05} consist of a
nematic  liquid crystal confined between a chemically homogeneous
grating surface (see Fig.~\ref{fig1}~(c)) and a flat substrate with
strong homeotropic anchoring. The profile of the asymmetric surface
grating is given by
\begin{align}
\label{eq:blazed}
z_0(x) = A \sin \left(qx+h\sin(qx)\right),
\end{align}
where $A$ is the groove depth, $p=2\pi/q$ the period, and
$h$ is the ``blazing'' parameter describing the asymmetry of the surface
profile. Such a grating surface has been studied by Brown {\it et al.}
\cite{brow:00} who found a first order transition between the HAN state,
characterized by a low pretilt angle ($\tilde \theta_0 \approx \pi/2$), and
the H state, characterized by a high pretilt angle ($\tilde \theta_0
\approx 0$). Strictly speaking, the H state does not correspond to the 
homeotropic texture (see Fig.~\ref{fig5}~(b) below),
but we keep the same notation as in the previous section for
consistency. Here we study phase transitions of a 
nematic liquid crystal in contact with the blazed surface in a more detail 
using the effective free energy method discussed in Sec.~II~C.

\begin{figure}[t!]
\includegraphics[width=8cm]{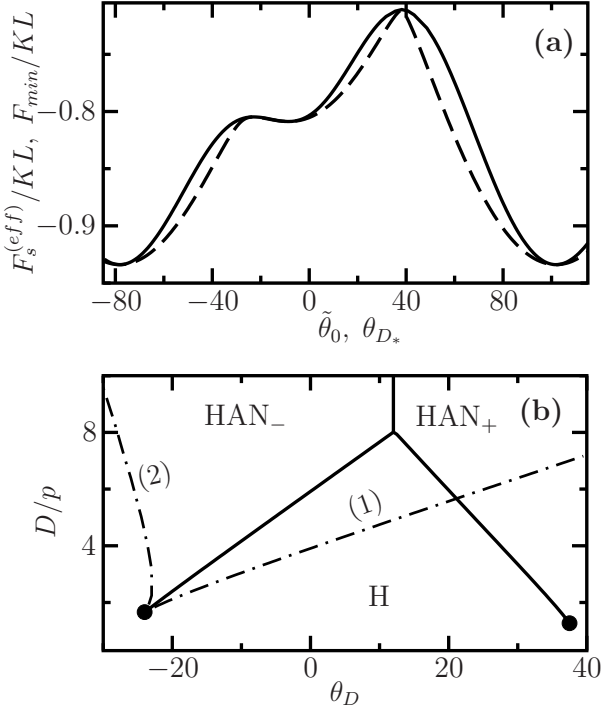} 
\caption{(a) The minimized free energy \mbox{$F(\theta_{D_*},D_*=p)|_{min}$} (dashed line) 
and the effective surface free energy
\mbox{$F_s^{(eff)}(\tilde{\theta}_0,D_*=p)$} (solid line) of a nematic
liquid crystal confined between a flat surface with strong anchoring 
and a chemically uniform blazed surface (see Fig.~\ref{fig1}~(c) and
Eq.~(\ref{eq:blazed})). The locally homeotropic anchoring strength on the blazed
surface is $pw_H/K=2$, the groove depth is $A/p=0.27$ and the blazing
parameter is $h=0.2$. $L$ is the extension of the cell in the invariant 
$y$ direction and $K$ is the isotropic elastic constant. The
effective surface free energy \mbox{$F_s^{(eff)}$} (as well as
\mbox{$F(\theta_{D_*},D_*=p)|_{min}$}) is periodic with the period
$2\pi$ but it is asymmetric with respect to $\tilde{\theta}_0=0$. 
(b) Phase diagram of the same system
as a function of the anchoring angle at the upper flat substrate
$\theta_D$ and the cell width $D$ with the
same line code as in Figs.~\ref{fig2} and \ref{fig3}. The triple point
(where the HAN$_+$, HAN$_-$ and H phases coexist) is at $\theta_D
\approx 12^\circ$ and $D/p \approx 8$, and the critical points (solid circle)
are at $D_{cr}/p \approx 1.3$, $\theta_D^{(cr)} \approx 38^\circ$ and
$D_{cr}/p \approx 1.7$, $\theta_D^{(cr)} \approx -24^\circ$. For
clearness, only the limits of metastability of the HAN$_-$ (1) phase and
the H (2) phase (for negative $\theta_D$) are shown.}
\label{fig:blazed}
\end{figure}

In Fig.~\ref{fig:blazed}~(a) the minimized free energy
$F(\theta_{D_*},D_*=p)|_{min}$ (dashed line) and the calculated
effective surface free energy $F_s^{(eff)}(\tilde{\theta}_0,D_*=p)$
(solid line) are shown for the anchoring strength $p w_H/K=2$
on the grating surface, the groove depth $A/p = 0.27$, and the blazing 
parameter $h=0.2$. The effective surface free energy is asymmetric 
with respect to ${\tilde \theta}_0=0$ because of the asymmetry of
the grating surface. As a consequence also the phase diagram, plotted as
a function of the anchoring angle on the upper surface $\theta_D$ and
the distance $D/p$ is asymmetric (Fig.~\ref{fig:blazed}~(b)). However, 
the topology of the phase diagram is the same as in the case of a symmetric 
substrate (see Figs.~\ref{fig2} (b), \ref{fig3} (b), and for a more 
general discussion Sec.~III~C below).

\begin{figure}[!t]
\includegraphics[width=8cm]{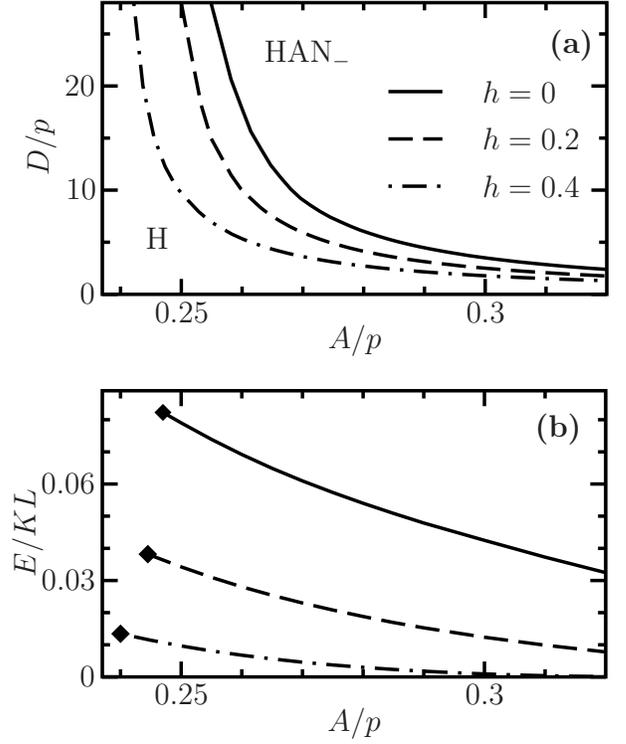} 
\caption{(a) The phase diagram of a nematic liquid crystal confined
between the blazed surface (see Fig.~\ref{fig1}~(c) and Eq.~(\ref{eq:blazed})) 
with local homeotropic anchoring and a flat surface with strong homeotropic
anchoring ($\theta_D=0$) as a function of the groove depth $A/p$ and the
cell width $D/p$. The anchoring strength on the blazed surface is
$pw_H/K=2$. The lines correspond to different values of $h$ and denote
first order transitions between homeotropic (H) and
hybrid aligned (HAN$_-$) phases. For small $A/p$, the lines extend to
$D=\infty$ corresponding to a first order {\it anchoring} transition
between planar and homeotropic phases. Upon increasing $A/p$ 
the first order transition lines end at critical points which are not 
shown in the figure. (b) The energy barrier at the first order transitions 
with the same line code as in (a). The lines have been obtained from the total
effective free energy (see Eq.~(\ref{eq6})), while the diamonds
correspond to the energy barriers between the planar and homeotropic
effective anchoring which follow from considering only the surface
contribution $F_s^{(eff)}$.}
\label{fig4}
\end{figure}
\begin{figure}[!t]
\includegraphics[width=8cm]{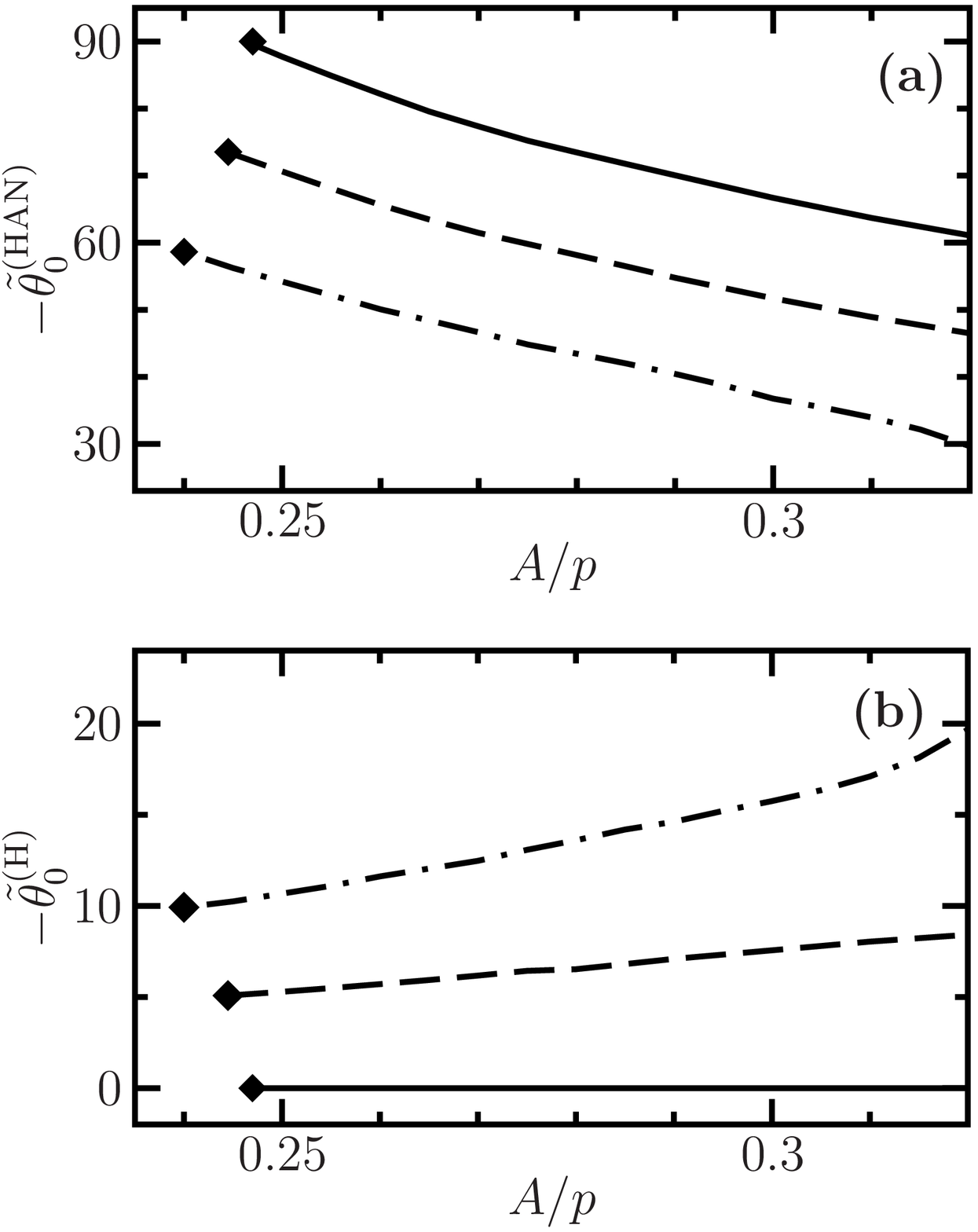} 
\caption{
The average surface director (see Eq.~(\ref{eq7})) for the hybrid
aligned (HAN$_-$) phase in (a) and the homeotropic (H) phase in (b) on the
lines of the first order transitions (see Fig.~\ref{fig4}) for a nematic
liquid crystal confined between the blazed surface (see Fig.~\ref{fig1}~(c) 
and Eq.~(\ref{eq:blazed})) with local homeotropic anchoring and a flat
surface with strong homeotropic anchoring ($\theta_D=0$). The diamonds
in (a) and (b) correspond  to the two minima of the effective surface
energy function $F_s^{(eff)}(\tilde
\theta_0)$. The model parameters and the line code are the same as in
Fig.~\ref{fig4}. 
}
\label{fig5}
\end{figure}

We now concentrate on the most interesting (from a practical point of
view) case of strong homeotropic anchoring ($\theta_D = 0$) at the upper
homogeneous surface. Figure \ref{fig4} (a) displays the phase diagram
for a few values of the blazing parameter $h$ and a fixed value of the
local homeotropic anchoring strength on the grating surface $pw_H/K=2$.
For a fixed cell width $D/p$, asymmetry ($h\ne 0$) leads to a decrease 
of the groove depth $A/p$ at which there is a first order transition 
between the HAN and the H phases, as compared to the nematic liquid 
crystal cell with the symmetric surface grating ($h=0$). Upon increasing 
the groove depth $A/p$ the transition line ends at a critical point 
(not shown in the figure), while it diverges as $A\to A_0$. The groove
depth $A_0$ corresponds to an {\it anchoring} (or surface) transition
between low tilt and high tilt {\it surface} states which are the
homeotropic and planar effective anchoring states, respectively, in the case $h=0$.

The effective free energy method allows one to calculate
an energy barrier $E$ between the two bistable states, which is
{\it not} feasible by the direct numerical minimization of the free
energy functional. The results of our calculations are shown in
Fig.~\ref{fig4}~(b). The asymmetry of the surface grating leads to a
decrease of the energy barrier. With increasing groove depth $A/p$
the energy barrier decreases and eventually vanishes upon
approaching the critical point. The diamonds in Fig.~\ref{fig4}~(b)
denote the energy barriers for the abovementioned anchoring transitions
of a nematic liquid crystal in contact with a single grating surface.

The energy barrier between two bistable states is an important quantity
for the design of a zenithally bistable nematic device. Too small energy
barrier, as compared to $k_BT$, would cause spontaneous switching
between the two states because of thermal fluctuations, while enlarging
the energy barrier leads to an increase of the power consumption. Using
the calculated values of $E$ (see Fig.~\ref{fig4}~(b)) one can estimate the 
energy barrier in a real nematic liquid crystal cell. For instance, for
a cell of area $1\,\mu \mathrm{m}\times 1\, \mu \mathrm{m}$ and of 
width $D = 5\, \mu \mathrm{m}$, and taking the typical  values $K =
5\times 10^{-12}\,\mathrm{N}$ and $w_H= 10^{-5} \,\mathrm{N/m}$, one
obtains $E \approx 37\, k_B T$ for $h = 0.2$ and $A \approx 0.026\,\mu
\mathrm{m}$, which seems to be an acceptable value.

Another important quantity in zenithally bistable nematic devices is the
average director orientation at the grating surface in the two
degenerate states. The average surface director in the HAN ($\tilde
\theta_0^{(\mathrm{HAN})}$) and H ($\tilde \theta_0^{(\mathrm{H})}$)
states is shown in Fig.~\ref{fig5} for the same model parameters as in
Fig.~\ref{fig4} and for the values of $A/p$ and $D/p$ corresponding to
the coexistence line. The asymmetry of
the surface grating leads to a decrease of $\tilde
\theta_0^{(\mathrm{HAN})}$ and an increase of $\tilde
\theta_0^{(\mathrm{H})}$. For a fixed value of $h$, the difference
between the two angles decreases with increasing the groove depth $A/p$
and finally vanishes upon approaching the critical point (not shown in
the figure).

Hence the asymmetry of the surface grating
leads to a decrease of the groove depth at which the bistability is
observed, which improves optical properties \cite{krie:02}, and to a
decrease of the energy barrier, which lowers the power consumption  of a
zenithally bistable nematic device. On the other hand, also the
difference between the two bistable states decreases which impairs 
optical properties of such a device.

\subsection{Phase diagrams for a model surface free energy}

  In section \ref{IIIA} we have discussed phase diagrams in the 
$(\theta_D,D/p)$ plane which can be described in terms of the surface
free energy given by Eq.~\eqref{eq14}. However, it is instructive 
to consider a more general situation that the surface 
free energy follows from a truncation of the Fourier expansion
given in Eq.~\eqref{eq18}. To be able to study both symmetric and 
asymmetric surfaces we assume a natural generalization of Eq.~\eqref{eq14}, 
namely
\begin{align}
\label{eq20}
   F_s(\theta_0) = \sum_{n=0}^{2}
             \left[ a_n \cos(2n\theta_0) + b_n \sin(2n\theta_0)\right]\,,
\end{align}
which reduces to Eq.~\eqref{eq14} in the case of a symmetric surface 
characterized by $b_n=0$. 
The angle $\theta_0^{(min)}$ that minimizes $F^{(eff)}(\theta_0;\theta_D,D)$
(see Eq.~\eqref{eq13}) is a function of $\theta_D$ and $D$. The derivative
\begin{align}
\label{eq21}
    \chi = \left(\frac{\partial\theta_0^{(min)}}{\partial\theta_D}\right)_D = 
    \frac{KLp/D}{KLp/D + F_s''(\theta_0^{(min)})} ,
\end{align}  
is the susceptibility of the system that diverges at the critical thickness
\begin{align}
\label{eq22}
       D_{cr} = - KLp/\min_{\theta_0} F_s''(\theta_0) ,
\end{align}
and remains finite and positive for $D<D_{cr}$. From Eq.~\eqref{eq22} the
conditions for the critical angle $\theta_0^{(cr)}$ follow as
\begin{align}
\label{eq23}
      F_s^{(3)}(\theta_0^{(cr)}) = 0 \;\;\;\textrm{and}\;\;\;
         F_s^{(4)}(\theta_0^{(cr)})>0\,,
\end{align}      
implying that $D_{cr}$ and $\theta_0^{(cr)}$ depend only on the form of $F_s(\theta_0)$.

\begin{figure}[t!]
\includegraphics[width=0.49\textwidth]{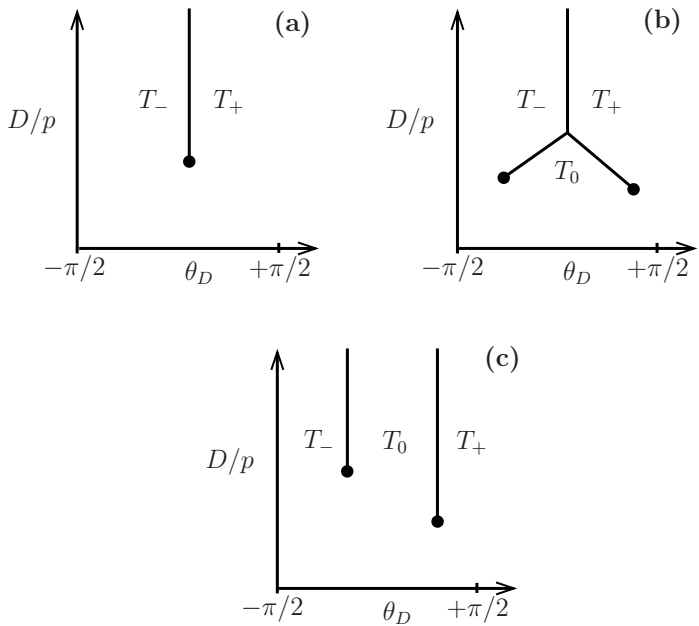}
\caption{
Schematic presentation of possible types of the phase diagram in the
$(\theta_D,D/p)$ plane for the surface free energy given by Eq.
\eqref{eq20}. $T_0$ and $T_{\pm}$ denote different  (usually
non-uniform) textures, the thick lines correspond to first order
transitions, and black circles  mark critical points. When $F_s$
(defined on the unit circle $z={\rm e}^{2i\theta_0}$) has one minimum
and one maximum the phase diagram can be either of type (a) or (b). When
$F_s$ has two minima and two maxima  the phase diagram is of type (b)
unless the minima are of equal depth, in which case it is of type (c).
Note that the lines $\theta_D=\pm\pi/2$ are identified with each other
and the phase diagram can be considered as being on a cylindrical
surface.}  
\label{fig:diagram}
\end{figure}

The extremes of $F_s$ given by Eq. \eqref{eq20} can be found easily only
in the case of symmetric or antisymmetric ($a_n=0$) surface and  the
same concerns  the position of critical point. In this work, however, we
are interested rather in possible topologies of the phase diagram in
the $(\theta_D,D/p)$ plane, which result from  Eqs. \eqref{eq13} and
\eqref{eq20}, and not in the exact location of critical points or
transition lines. To draw schematic phase diagrams we consider $F_s$ as
a function defined on the unit circle $z={\rm e}^{2i\theta_0}$.
Depending on the parameters $a_n$ and $b_n$, $F_s(\theta_0)$ has either
one minimum and one maximum or two minima and two maxima. This
conclusion applies also to the function $F_s''(\theta_0)$, thus, there
can be either one or two critical points  in the phase diagram (see Eqs.
\eqref{eq22} and \eqref{eq23}). In the limit of large $D$, there is
always a first order phase transition between two non-uniform textures
corresponding to the opposite orientations of the director at $z=0$.
This is because $F^{(eff)}$ is not  a periodic function of $\theta_0$ at
fixed $\theta_D$. To find $\theta_D$ at the transition we expand $F_s$
around its deepest minimum  (denoted $\theta_m$), which leads to the
approximate free energy:
\begin{align}
\label{Feq}
    F^{(eff)}(\theta_D,D)|_{min} \approx F_s(\theta_m) 
       + \frac{KLp(\theta_D-\theta_m)^2}{2(D+b)} ,
\end{align}
where $b=KLp/F_s''(\theta_m)$ is the extrapolation length. Since
$\theta_m$ and $\theta_m\pm\pi$ are equivalent  minima of $F_s$, and
both $\theta_m$ and $\theta_D$ are allowed  to vary in the interval
$[-\pi/2,\pi/2]$, the transition occurs at
$\theta_D^{(tr)}=\theta_m+\pi/2$ if $-\pi/2\leq\theta_m\leq 0$ or at
$\theta_D^{(tr)}=\theta_m-\pi/2$ if $0\leq\theta_m\leq\pi/2$.  With the
above information we can now draw schematically the phase diagram (see
Fig.~\ref{fig:diagram}).  Since the vertical lines at
$\theta_D=\pm\pi/2$ are to be identified with each other the phase
diagram can be considered on a cylindrical surface. Away from the
transition lines there is  a smooth evolution from one texture to
another. This means that at fixed $D$ it is possible to transform
smoothly the $T_{+}$ texture into the $T_{-}$ texture, i.e., without
crossing the transition line, even for $D>D_{cr}$. We note that in some
range of parameters, the phase diagram for $F_s$ with one minimum is
topologically indistinguishable from that for $F_s$ with two minima
(Fig.~\ref{fig:diagram}~(b)); in both cases there are two critical points 
and a triple point. If $F_s$ has two equal minima (e.g., at
$\theta_0=0$ and $\theta_0=\pm\pi/2$ in the case of symmetric surface)
the triple point disappears and the first order transition lines extend
to $D\to \infty$,  as shown in Fig.~\ref{fig:diagram}~(c).

\section{Summary}

We have studied the phase behavior of a nematic liquid crystal confined
between a flat and a patterned substrate (Fig.~\ref{fig1}) using the
Frank-Oseen model  [Eq.~(\ref{eq4})] and the Rapini-Papoular surface
free energy [Eq.~(\ref{eq5})].  An expression for the effective free
energy function of the system [Eq.~(\ref{eq6})]  was derived by
determining an effective surface free energy characterizing the 
anchoring energy at the patterned surface [Eq.~(\ref{eq8})]. Using the
effective  free energy function, we have determined the phase behavior
of the nematic liquid crystal confined between a flat surface with 
strong anchoring and a chemically patterned sinusoidal surface 
(Fig.~\ref{fig1} (a)), finding first order transitions between
a homeotropic texture (H) and hybrid aligned nematic (HAN) textures
(Figs.~\ref{fig2} (b) and  \ref{fig3} (b)). It is  possible to have a
first order phase transition even with a monostable surface
characterized by a monotonic surface free energy function
(Fig.~\ref{fig2}~(a), ${\tilde \theta}_0\in [0,\pi/2] $). 
In addition we have performed direct
minimizations of the original free energy functional  [Eqs.~(\ref{eq4})
and (\ref{eq5})] on a two-dimensional grid and found remarkably  good
agreement with the phase boundaries resulting from the effective energy 
function analysis (Figs.~\ref{fig2} (b) and  \ref{fig3} (b)).  Hence
quantitatively reliable predictions of the phase behavior can be
achieved using the effective free energy method.

Using this method, we have also studied the phase behavior 
(Fig.~\ref{fig:blazed} (b)) of a nematic liquid crystal confined between a 
chemically uniform, asymmetrically grooved substrate (Fig.~\ref{fig1}~(c) 
and Eq.~(\ref{eq:blazed})) with locally homeotropic anchoring and a flat 
substrate with strong homeotropic anchoring, which is a typical setup for 
a zenithally bistable nematic device \cite{parr:03,brow:04,parr:04,parr:05}.
The asymmetry of the grating substrate leads to a decrease of the groove
depth at which a first order transition between the H and HAN phases
occurs (Fig.~\ref{fig4}~(a)). Moreover, we have determined the energy barrier
between the two coexisting states (Fig.~\ref{fig4}~(b)). Our calculations show 
that the energy barrier decreases with increasing the
asymmetry of the grating surface but it is well above $k_B T$ for a 
typical nematic liquid crystal cell. In addition, the average director 
orientation at the grating surface in two bistable states has been calculated
(Fig.~\ref{fig5}). The difference between the two bistable
states vanishes with increasing substrate asymmetry, which has
a negative effect on the optical properties of a zenithally bistable 
nematic device.

We have also generalized the model of the effective surface free energy 
considered by Parry-Jones {\it et al.} \cite{parr:03,parr:04} to the
case of asymmetric structured substrates and obtained three possible
types of the phase diagram in the plane spanned by the orientation of
the director at the homogeneous surface and the thickness of the nematic cell. 
The asymmetry of the substrate causes only a shift of transition lines and 
critical points, compared to the symmetric case, but does not change the 
topology of the phase diagram. Finally, we have verified that this model 
allows one to reproduce qualitatively the phase diagram of a nematic 
liquid crystal confined between a homogeneous planar substrate and an 
asymmetrically grooved surface (Fig.~\ref{fig:blazed}~(b) and Fig.~\ref{fig:diagram}~(b)).

\end{document}